\documentclass[manuscript]{aastex}

\usepackage[utf8]{inputenc}
\usepackage[intlimits]{amsmath}
\usepackage{amssymb}
\usepackage[amssymb]{SIunits}
\usepackage{graphicx}
\usepackage[dvipdfm]{hyperref}
\usepackage{natbib}
\usepackage[normalem]{ulem}

\newcommand{\rv}[1]{#1}
\newcommand{\rvd}[1]{}
\newcommand{\rvi}[1]{#1}
\newcommand{\rvdi}[1]{}

\newcommand{\aas}{Astron. Astrophys. Suppl. Ser.}

\DeclareMathOperator{\sech}{sech}
\DeclareMathOperator{\stdev}{stdev}
\newcommand{\mean}[1]{\overline{#1}}
\newcommand{\ud}{\mathrm{d}}

\newcommand{\Msun}{\text{M}_\odot}
\newcommand{\AU}{\text{AU}}
\newcommand{\erg}{\text{erg}}
\newcommand{\parsec}{\text{pc}}
\newcommand{\mjy}{\text{mJy}}
\newcommand{\yr}{\text{yr}}
\newcommand{\magn}{\text{mag}}
\newcommand{\risco}{r_\text{ISCO}}
\newcommand{\rsch}{r_\text{sch}}

\newcommand{\rsec}{r_\text{sec}}
\newcommand{\medd}{\dot{M}_\text{Edd}}
\newcommand{\medds}{\dot{m}_\text{Edd}}
\newcommand{\Ledd}{L_\text{Edd}}
\newcommand{\vrel}{v_\text{rel}}

\newcommand{\prim}{1.84\times 10^{10}\usk\Msun}
\newcommand{\secm}{1.4\times 10^8\usk\Msun}

\newcommand{\consthsymb}{z_c}
\newcommand{\consth}{4000\usk\AU}

\newcommand{\pci}{1993.93}
\newcommand{\pcii}{2004.27}
\newcommand{\pciii}{2012.29}

\newcommand{\nextpc}{2020.96}
\newcommand{\nextpcerr}{0.10}
\newcommand{\nextb}{2022}

\begin{document}

\title{Precursor flares in OJ 287}

\author{P. Pihajoki\altaffilmark{1},
M. Valtonen\altaffilmark{2},
S. Zola\altaffilmark{3,4},
A. Liakos\altaffilmark{5}, 
M. Drozdz\altaffilmark{4}, 
M. Winiarski\altaffilmark{4}, 
W. Ogloza\altaffilmark{4}, 
D. Koziel-Wierzbowska\altaffilmark{3}, 
J. Provencal\altaffilmark{6,7},
K. Nilsson\altaffilmark{2},
A. Berdyugin\altaffilmark{1},
E. Lindfors\altaffilmark{1},
R. Reinthal\altaffilmark{1},
A. Sillanpää\altaffilmark{1},
L. Takalo\altaffilmark{1},
M.M.M. Santangelo\altaffilmark{8,9},
H. Salo\altaffilmark{10},
S. Chandra\altaffilmark{11},
S. Ganesh\altaffilmark{11},
K.S. Baliyan\altaffilmark{11},
S.A. Coggins-Hill\altaffilmark{12},
and
A. Gopakumar\altaffilmark{13}
}
\email{popiha@utu.fi}

\altaffiltext{1}{Tuorla Observatory, Department of Physics and Astronomy, University of Turku, 21500 Piikkiö, Finland}
\altaffiltext{2}{Finnish Centre for Astronomy with ESO, University of Turku, 21500 Piikkiö, Finland}
\altaffiltext{3}{Astronomical Observatory, Jagiellonian University, ul.  Orla 171, 30-244 Krakow, Poland}
\altaffiltext{4}{Mt. Suhora Observatory, Pedagogical University, ul. Podchorazych 2, 30-084 Krakow, Poland}
\altaffiltext{5}{Department of Astrophysics, Astronomy and Mechanics, University of Athens, GR 157 84 Zografos, Athens, Hellas, Greece}
\altaffiltext{6}{Department of Physics and Astronomy, University of Delaware, Newark, DE 19716, USA} 
\altaffiltext{7}{Delaware Asteroseismic Research Center, Mt. Cuba Observatory, Greenville, DE 19807, USA}
\altaffiltext{8}{O.A.C. Osservatorio Astronomico di Capannori, Via di Valle, 55060 Vorno, Capannori, Italy}
\altaffiltext{9}{I.R.F. Istituto Ricerche Fotometriche, Viale Luporini trav.2 N.111, 55100 Lucca, Italy}
\altaffiltext{10}{Department of Physical Sciences, University of Oulu, P.O. Box 3000, 90014 University of Oulu, Finland}
\altaffiltext{11}{Astronomy \& Astrophysics Division, Physical Research Laboratory, Ahmedabad 380009, India}
\altaffiltext{12}{Am Weinberg 16, 63579 Freigericht-Horbach, Germany}
\altaffiltext{13}{Tata Institute of Fundamental Research, Mumbai 400 005, India}

\begin{abstract}
We have studied three most recent precursor flares in the light curve of the
blazar OJ 287 while invoking the presence of a precessing binary black hole in
the system to explain the nature of these flares. Precursor flare timings from
the historical light curves are compared with theoretical predictions from our
model that incorporate effects of an accretion disk and post-Newtonian
description for the binary black hole orbit. We find that the precursor flares
coincide with the secondary black hole descending towards the accretion disk of
the primary black hole from the observed side, with a mean $z$-component of
approximately $\consthsymb = \consth$.  We use this model of precursor flares
to predict that precursor flare of similar nature should happen around
$\nextpc$ before the next major outburst in $\nextb$.

\end{abstract}


\maketitle


\section{Introduction}

\object{OJ 287} is a blazar at redshift $z=0.306$ 
that exhibits nearly periodic double peaked outbursts
at intervals of approximately 12 years in the optical regime
\citep{sil88,sil96a,sil96b}.
This periodicity is not exact, as was definitely demonstrated by the observed
outbursts in 2005 and 2007 \citep{val06a,val08a,val08b}.
Both the double peaked structure and the dwell in the outburst interval
have been successfully explained by a precessing binary black hole
model \citep{leh96,sun96,sun97,val11}.  
In this model, the double optical outbursts are caused by a secondary black hole
impacting the accretion disk of the primary black hole, twice in one
period.

The orbit of the secondary is sufficiently compact and eccentric ($e\sim
0.7$) to bring it close enough to the primary for strong relativistic
precession of the orbit \citep{val97,val07}.
The magnitude of the precession has been established at approximately
$39.1\degree$ per cycle \citep{val10a}. 
The binary black hole with orbital speeds close to
$10 \%$ of the light speed enabled probing the conservative aspects of
general relativistic binary dynamics 
at the second post-Newtonian order, including the effects of dominant order
spin-orbit coupling \citep{val10a}.
Further, the possibility of testing in the near future 
general relativity and black hole properties using
the binary black hole in OJ 287 makes the detailed understanding of 
various observational features of the blazar especially important \citep{val11}.

\begin{figure*}[htpb]
\center{\includegraphics*{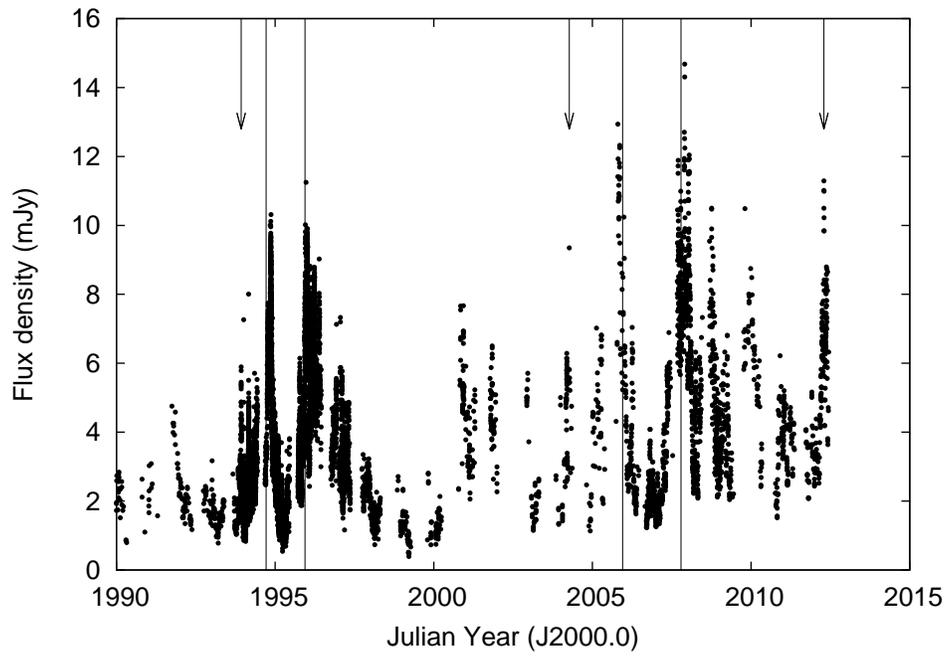}}
\caption{Recent light curve of the OJ 287 
in V-band equivalent flux. Precursor timings are marked with arrows.
Calculated major optical outburst timings of the model in \cite{val10b} are
marked with vertical lines.}
\label{fig:lightcurve}
\end{figure*}

In this paper we discuss an aspect of OJ 287 binary black hole which has
attracted rather little attention so far. The repeated impacts of the
secondary black hole on the accretion disk of the primary create a
corona of high velocity gas clouds above the standard geometrically thin
but optically thick accretion disk. As the secondary travels through
this disk, it may accrete some of the clouds, and these accretion events
may lead to quite prominent brightening of the OJ 287 system as whole.
\rvd{
They would be akin to accretions of whole stars by a black hole, as has
been recently reported \mbox{\citep{gez12}}. 
}
\rv{They could be contrasted with similarly fast brightening events of
black holes accreting whole stars, as has been recently reported
\citep{gez12}, but with a different timescale and mechanism of action.}
If such events can be identified,
they would allow the detection of the secondary black hole, in addition
to the primary which is normally responsible for the emission of OJ 287. 

The type of light curve features we will discuss are shown in
Figure~\ref{fig:lightcurve} which
displays the optical light curve of OJ 287 since 1990. Besides the double peak
structure, indicated by the vertical lines, there are prominent outbursts prior
to each cycle called \emph{precursors} by \cite{kid93}. They are marked by
arrows in Figure~\ref{fig:lightcurve}. The first one peaked at $\pci$ and it was studied by
\cite{kid94,kid95}. The second one peaking at $\pcii$  was studied by
\cite{val06b}. They also discussed a possible origin of the precursors.
The third precursor occurred only recently, peaking at $\pciii$, and the
data related to this precursor are reported here. We will also describe
two possible scenarios with regard to the origin of precursors and
finally settle on a model that seems to fit the limited amount of data.

\section{Observations}

In addition to the previously published light curve 
\citep{val09}, 
we have
added new datapoints up to the most recent precursor flare. These observations
have been made at 
Tuorla Observatory in Finland (hereafter labelled TUO),
Astronomical Observatory of Capannori in Italy (OAC), 
Astronomical Observatory of the Jagiellonian University (KRK) 
and the Mt.  Suhora Observatory of the Pedagogical University (SUH) in Poland, 
University of Athens (ATH) in Greece,
Mt. Cuba Observatory (MTC) in USA,
Mt. Abu Infrared Observatory (MIRO) in India
and
Liverpool Telescope (LIV) and Kungliga Vetenskapliga Akademien Telescope (KVA) 
situated at Observatorio del Roque de Los
Muchachos in La Palma, Canary Islands, Spain.

\begin{figure}[htpb]
\center{\includegraphics{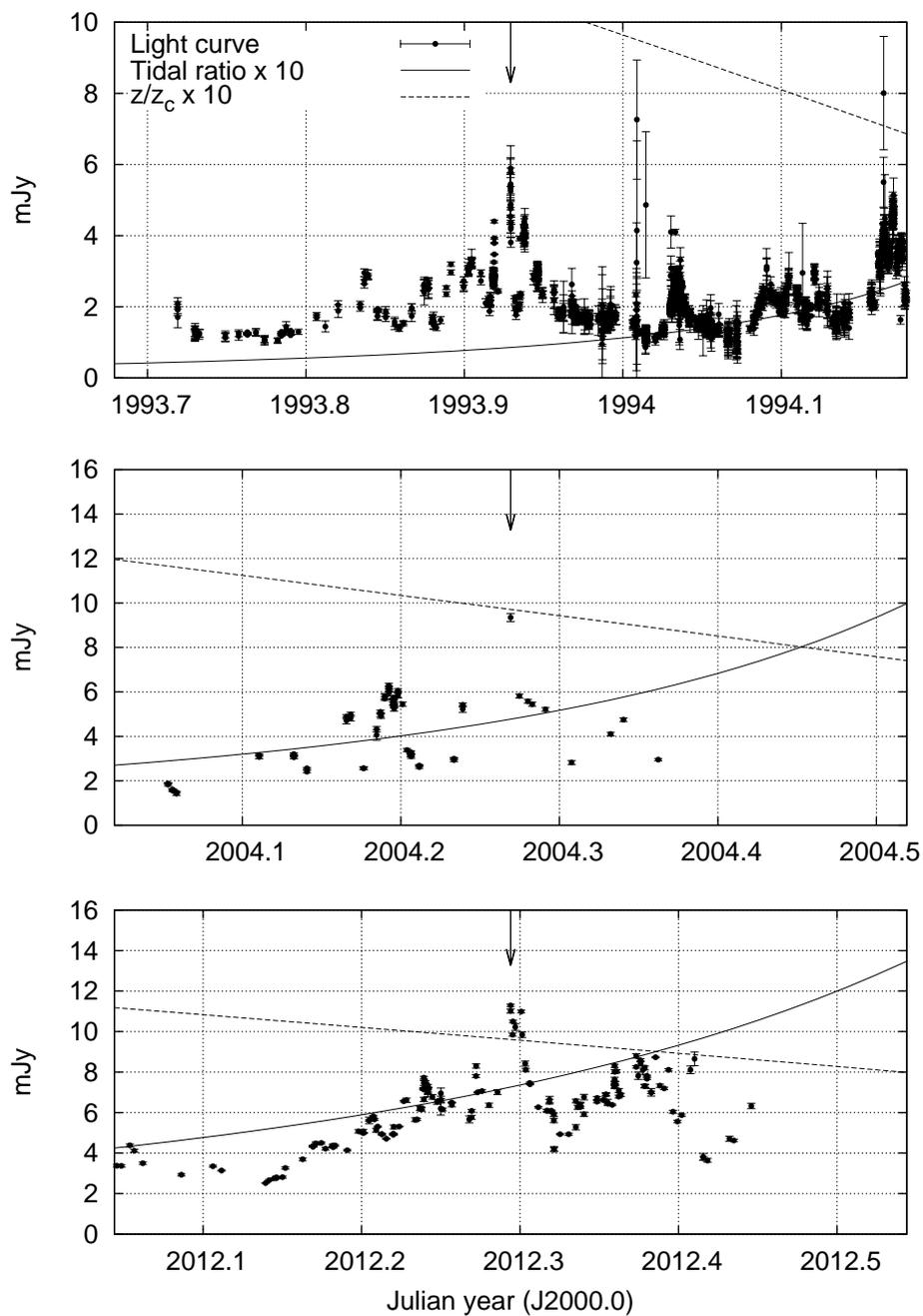}}
\caption{Light curve with error bars near the precursor flares. 
Flare timings are marked with arrows. The solid line gives
$10\times$ the tidal ratio while the dashed line shows $10\times$
the height of the secondary above the accretion disk in units of
$4084\usk\AU$.}
\label{fig:closeups}
\end{figure}

The new measurements are catalogued in Table~\ref{tb:observations}. They
have been carried out in V, B and R bands, converted to V-magnitudes and
then to V-band fluxes (for details, see e.g. \cite{val08b}). These light
curve points are shown with greater resolution in the lowest panel of
Figure~\ref{fig:closeups}. The timing of the peak flux is at $\pciii$. The two other
panels of Figure~\ref{fig:closeups} display the corresponding times around earlier
precursors at the same temporal resolution.

We may note that typically there is more than one precursor peak. We
will generally discuss the timing of the highest peak, but actual
observed maximum height depends on how frequently the observations have
been made. The precursors are very rapid phenomena in comparison with
the standard double peak outbursts, and therefore the times of maxima
may easily pass unobserved. 
\rv{A qualitative definition of a precursor
flare can be given as intense flaring activity just before a scheduled
major outburst, not exceeding it in brightness.
To objectively classify the precursor peaks we note the following:
The twin major outbursts typically rise $2.5\usk\magn$ or more in V-band
brightness above the brightness in quiescent state. This quiescent
brigtness level can be defined as the level under which the V-band
brightness doesn't fall, with measurement errors taken into account.
This level has been rising gradually; in 1990--1994 it was $\sim
1\usk\mjy$, in 2000--2005 $\sim 1.4\usk\mjy$ and from 2010 onwards
$\sim 1.8\usk\mjy$.
For the precursors, we then have a defining limit of a rise of
$2\usk\magn$ or more in V-band compared to quiescent level.
We note that objective classification of precursor flares is not
entirely straightforward, as OJ 287 exhibits natural flaring activity
induced by the tidal action on the primary accretion by the close passes
of the secondary.
}
For the following discussion it is not
important which choice of the precursor peak \rv{matching the criterion} 
we make.

\begin{deluxetable}{ccccc}
\tablecaption{\label{tb:observations}Observations of the 2012 precursor
flare.}
\tablehead{\colhead{MJD} & \colhead{Location} & 
\colhead{Band} & \colhead{Mag.} & \colhead{Mag. error}}
\startdata
55112.2675400 & KVA & R & 14.107 & 0.027 \\
55113.2541600 & KVA & R & 13.921 & 0.019 \\
55118.2574200 & KVA & R & 14.089 & 0.018 \\
55120.2617300 & KVA & R & 13.993 & 0.018 \\
55126.2527500 & KVA & R & 13.468 & 0.018 \\
\enddata
\tablecomments{Table 1. is published in its entirety in the electronic
edition of The Astrophysical Journal. A portion is shown here for
guidance regarding its form and content.}
\end{deluxetable}

\subsection{Observations at TUO and KVA}

The observations were made with the Tuorla 1.03 m telescope
at Tuorla Observatory, Finland and with the 35 cm KVA telescope on
La Palma, Canary Islands through the R-band filter. A single observation
typically consisted of 3-4 exposures of 60s (Tuorla) or 180s (KVA).
The data were reduced in the usual way by first subtracting the bias
and dark frames and then dividing by a dome flat-field. The brightness
of OJ 287 and star 4 with $R = 13.74$ \citep{fio96} 
were measured with aperture photometry after which the R-band
magnitude of OJ 287 was derived from the measured magnitude difference
using color corrections appropriate for the two telescopes.

\subsection{Observations at SUH, KRK, ATH and MTC}

OJ 287 has been frequently monitored in the R band 
filter at SUH, KRK and ATH since 2006. Additional data
in the R filter has been gathered at MTC in
the middle of the 2012 precursor outburst with the 
60 cm telescope and an Apogee Alta U47 CCD mounted at
the prime focus.
The 60 cm telescope equipped also with an Alta U47 CCD 
installed at the telescope primary focus was used at SUH 
while the 50 cm telescope and an Andor iKon-L DZ936N-BV
camera at KRK.  
The data at ATH were gathered with the 40 cm Mead 
telescope and a SBIG ST-10 CCD. Cameras at ATH and KRK
are mounted at the Cassegrain foci of the telescopes.
We took bias and dark calibration images every night,
and whenever possible, sky flatfield images were taken,
otherwise, dome flats were used.
Each night we took 10--20 frames of OJ 287, usually in
the R filter, but on several occassions we measured
the blazar in UBVRI wide band filters. \emph{MaximDL} software
was used for data gathering at ATH while the \emph{JAstroCam}
program \citep{bud10}, developed 
to run under the Linux OS, was deployed at SUH, KRK and MTC 
observatories. 

The reduction of data taken at the four sites was done 
by one person to secure uniformity over long period of time. 
Images of OJ 287 have been corrected for bias, dark and flatfield 
making use of the \emph{IRAF} package and differential photometry 
was performed to extract magnitude differences.
The \emph{CMunipack} program\footnote{http://c-munipack.sourceforge.net},
which is an interface for the \emph{DAOPHOT} package, was applied, 
stars no. 4 and no. 10 \citep{fio96}  
served as comparison and check stars, respectively. 
All individual measurements from each night have been 
averaged and the standard deviations calculated.

\subsection{Observations at OAC}

A total of 89 BVRI (17 B, 22 V, 28 R, 22 I) (Johnson/Cousins) CCD frames
of OJ 287 were taken at OAC on 18 nights in Spring 2012, from JD
2456015 to JD 2456066, with the $0.30\usk\metre$ $f/10$ telescope equipped with 
Sbig ST-9XE CCD camera, Sbig CFW-9 filters wheel, Custom Scientific BVRI
glass filters, and Sbig AO-8 adaptive optics. On each night also bias
frames, master median dark frames, and master median flat field (with
both
twilight $+$ diffuser, and dome $+$ diffuser) frames were taken.  

On some 
of the nights used for OJ 287, BVRI frames were taken also of the
CCD secondary photometric sequence in the so-called dipper asterism
region in M67 \citep{anu94}.  On two of the nights used
for OJ 287, BVRI frames were taken also of the CCD primary photometric
sequence around PG1633+099 \citep{lan09}.  
The frames of M67 and PG
1633+099 sequences were taken in order to derive independent estimates
of the coefficients of the colour equations needed to transform
OAC's instrumental bvri system to the standard BVRI one; but these
frames were not used.  Instead, on each observing night, these
transformation coefficients were derived from stars of the secondary
photometric UBVRI photometric sequence, which is claimed to be in
Landolt's UBVRI system, given by \cite{gon01} for stars
just around OJ 287. 

BVRI CCD synthetic aperture photometry on the OJ
287 frames was performed at OAC, using the \emph{AIP4 for Win} data reduction
software. The stars 9 and 13 of the photometric sequence around OJ 287
established by \cite{gon01} were used respectively as
comparison and as check stars. The BVRI (Johnson/Cousins) magnitudes of
star 9 and star 13 were taken from table 8 of \cite{gon01}. The times
of these OAC's time series were corrected for the heliocentric
correction. From the final standard magnitude datapoints, those in B and
V bands were used in this publication.

\subsection{Observations at MIRO}

A total of 1608 exposures were taken in Johnson/Cousins R band on 12
nights in Spring of 2011 and 2012. The CCD used is a liquid nitrogen
cooled Pixellent 1296x1152 pixel EEV, Grad 0, with a pixel size of 22
microns and field of view of approx 6.5'$\times$5'. Details of the
telescope, observation methods and data reduction pipeline can be found
in \cite{bal05,cha11}.

\subsection{Observations at LIV}

A total of 298 observations of OJ 287 were taken using CCD frames over a total
of 34 nights from JD 2455479.727 to 2456015.357. The observations were taken
using the $2\usk\metre$ robotic Liverpool Telescope, equipped with the RATCam
optical CCD fitted with clear Sloan r', Sloan g' and Bessel B filters. The
number of exposures with each filter was:  97 r', 101 g', 100 Bessell B.
Aperture photometry was then performed on the OJ 287 frames using the
\emph{AstroArt 4.0} astronomical image processing software. Finally, the
frames were analysed using an Excel spreadsheet.

For each observation, instrumental and standard magnitudes for OJ 287 were
derived from two stars (star 10 and star 4) of the secondary photometric
sequence given in \cite{gon01}. Star 10 and star 4 were selected as the
comparison star and check star respectively.

The published BVRI (Johnson/Cousins) magnitudes of star 10 and star 4 were
taken from table 8 of \cite{gon01} and from these the transformed standard
magnitudes for the two stars in the Sloan and Bessel filters were derived using
transformation formulae in \cite{jes05} for stars with $R-I < 1.15$. 

From the final standard magnitude data points, those in the B band were
included in this paper.

\section{Possible theoretical explanations}

\begin{figure}[h!tbp]
\plotone{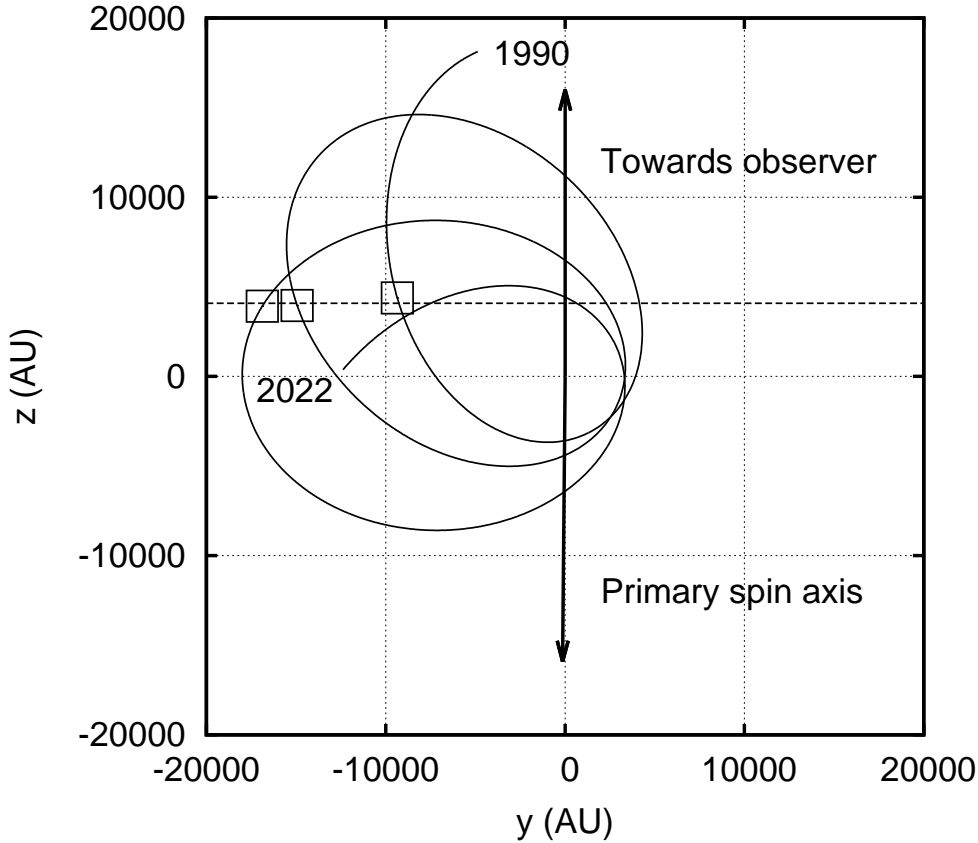}
\caption{Projection of the relative orbit of the secondary black hole to
$yz$-plane. Positions corresponding to precursor flare times are indicated by
boxes. The accretion disk of the primary is in the $xy$-plane. }
\label{fig:orbit_bursts}
\end{figure}

We have calculated the position of the secondary black hole relative to the
primary at the times of the precursor flares, using a Post-Newtonian integration
scheme of order 3, with binary orbital parameters derived from \cite{val10a}.
Speficifally, binary semimajor axis was $11507.957\usk\AU$, eccentricity
$0.658$ and initial precession angle at apocentre $56.3\degree$ at epoch
$1855.89679\usk\text{yr}$.
The results shown in Figure~\ref{fig:orbit_bursts} indicate that the precursor
bursts seem to occur when the secondary is at a constant height of
$\consthsymb\sim\consth$ above the $z=0$ plane of the accretion disk of the
primary. Thus we have to consider what mechanism would initiate outbursts at
this particular orbital phase. 

Let us first consider outbursts arising in the primary. They could be initiated
by perturbations in the accretion disk; these perturbations propagate to the
jet and initiate optical outbursts with about three months time delay
\citep{val06b}, a reasonable time interval in a strongly magnetic disk. Such
perturbations should start when the tidal perturbation of the secondary on the
accretion disk reaches some critical level.

To study this, we have calculated the ratio of the vertical tidal force
$F_2$,
caused by the secondary on an element of the accretion disk immediatelly
below it, to the horizontal tidal force $F_1$ caused by the primary on the
same disk element. 
The tidal force is proportional to the mass of the black hole and the cube of the distance
from it, and the ratio of the forces is thus
\begin{equation}
    \frac{F_2}{F_1} = \frac{m_2}{m_1}\frac{r^3}{z^3},
\end{equation}
where $m_i$ are the black hole masses, $r$ is the distance from the
primary in the plane its accretion disk, and $z$ is the vertical
distance of the secondary from the plane of the disk.
The tidal ratio (multiplied by 10) has been plotted as a
solid line in the three panels of Figure~\ref{fig:closeups}. 
We see that the tidal line describes
well the general rise of the activity level in OJ 287, and thus supports the
view that this background activity does arise from the primary black hole with
its associated disk and jet. However, there is no single threshold level where
the precursors start to arise. Thus this explanation does not seem to work for
the origin of precursors. It is possible that the tidal process is more
complicated than our simple tidal ratio estimate. We will carry out a full tidal
perturbation calculation in the next section to check this preliminary
conclusion.

The second possibility is that the source of the precursor peak is the secondary
black hole. In this scenario the secondary meets a layer of gas clouds at a
constant height above the disk, accretes these clouds in individual accretion
events, and this produces brightness spikes. To study this scenario, we have
plotted the height of the secondary above the accretion disk as a function of
time in the three panels of Figure~\ref{fig:closeups} (dashed line). The height
is measured in units of $\consthsymb=\consth$ (and multiplied by $10$ in the
display). We see that the constant height is a good predictor for the start of
the precursor events, as is also shown by Figure~\ref{fig:orbit_bursts}. 

This result leads us to consider a model where the precursor flares arise in
the secondary black hole when it plunges into a gas cloud in the
corona of the primary accretion disk. 
A natural origin for these clouds are the gas clouds pulled off by the
secondary from the accretion disk of the primary during previous orbital
revolutions. If the secondary impacts these clouds, some of the gas may
be accreted by the secondary, with subsequent brightening of its jet. 

We may assume the secondary black hole to be maximally spinning, i.e. with Kerr
parameter $\chi\sim1$ and with a mass of $\secm$ \citep{val10b}. This $\chi$
value is supported by the observed 228 minute periodicity \citep{sag04} which
is the period of innermost stable orbit for a black hole of this mass and spin.
It is also reasonable that the secondary has close to maximum spin as it
crosses the primary disk twice in 12 years and every time accretes matter with
the same direction of the specific angular momentum, the spin direction of the
accretion disk.

The spin of the primary has already been established at $\chi\sim0.28$
and its mass at $\prim$ \citep{val10b}. With these values the relative
electromagnetic brightness of the secondary versus the primary jet at
the times of the outbursts may be rather equal for comparable mass
accretion rates \citep{haw06,nei11}.

Looking at the next time the secondary orbit descends to the height of
$\consthsymb$ we can predict the timing of the next precursor flare.
\rvd{
It should occur around $\nextpc$, before the major outburst in $\nextb$.
However, as can be
seen from the Figure~\ref{fig:orbit_bursts}, the secondary barely rises above
the level $\consthsymb$ due to the strong orbital precession. This might affect
the accuracy of the prediction.
}
\rv{
The mean value and one sigma limits for $\consthsymb$ give
$\consthsymb = (4084 \pm 254)\usk\AU$. From this we find
the timing of the next expected precursor burst with one sigma limits to
be approximately $\nextpc\pm\nextpcerr$. This prediction leaves an ample
interval for observation before the next expected major outburst in $\nextb$.
}

\section{Numerical simulation}

To simulate the dynamics of the OJ 287 system, we have developed a
multiprocessor N-body solver along the lines of \cite{sun97}, but with some
significant improvements. The primary improvement is the simultaneous
evolution of black hole binary orbits and  accretion disk
particles around the primary black hole, including interactions between the
accretion disk particles themselves.

The accretion disk is modeled as a cloud of point-like particles initialized
with a standard vertical $\sech^2(z/z_0)$ distribution, where $z_0$ is the
vertical scale height. The point particles interact gravitationally with both
black holes in the binary and with each other through a grid-based viscosity
calculation.

For the disk viscosity calculation we use a radially nonuniform 
polar grid adapted from \cite{mil76}. 
In this scheme we define the grid
by coordinates $u$ and $v$ such that
\begin{subequations}
\begin{gather}
r(u) = R_c \exp(\alpha u) \\
\theta(v) = \alpha v \\
\intertext{where}
\alpha = \frac{2\pi}{N_\theta} \\
R_c = R_m\exp\left(-\frac{2\pi N_r}{N_\theta}\right).
\end{gather}
\end{subequations}
The parameter $R_m$ defines the maximum radial extent of the grid.
Parameters $N_r$ and $N_\theta$ then define the number of radial and
azimuthal divisions when
\begin{subequations}
\begin{gather}
0 \leq u < N_R \\
0 \leq v < N_\theta.
\end{gather}
\end{subequations}
The benefit of this formulation is that the grid cells are
approximately square through the entire radial extent of the grid.

The disk viscosity is calculated in the vertical and radial directions only,
using the physical model of \cite{leh96}.  The kinematic viscosity coefficient
$\nu$ of the disk gas as a function of the radial distance is obtained from
formula
\begin{equation}
\nu = A
\left(\frac{r}{\parsec}\right)^{3/2}
\left(\frac{M}{\Msun}\right)^{-1/2}
\left(\frac{n}{\centi\meter\cubed}\right)^{-1}
\left(\frac{T}{\kelvin}\right)^{4},
\end{equation}
where $r$ is the radial distance, $M$ is the central black hole mass, $n$ is the
particle density in the disk, and $T$ is the disk temperature. The last three
are functions of $r$ and are obtained from the model in \cite{leh96}. For
$M$ we use the primary black hole mass from \citep{val10b}.
Constant $A = 15914\usk\AU^2$ in a system of units defined by Solar mass, Julian
year and the gravitational constant $G=1$.
The velocity components $v_i$ of each particle in a
grid-cell are
compared to the mean velocity components $\mean{v_i}$ of all particles in the
cell. The viscosity force per unit mass is then calculated
from the component differences by the formula
\begin{equation}
f_i = -\alpha\nu\left(v_i-\mean{v_i}\right),
\end{equation}
in which $\alpha$ is the usual parameter in the $\alpha$-disk
theory \citep{sha73} and in its subsequent developments 
\citep{sak81}.

Additionally, the solver uses Post-Newtonian approximation scheme to
calculate the gravitational interaction of the black hole binary and the
forces between the black holes and the accretion disk particles.
In these numerical runs, binary dynamics is fully third post-Newtonian order accurate and the required formulas may 
be found in \cite{val10a}.

The model was used to calculate particle accretion counts to the primary and
secondary black holes. The accretion radii were set to $10\rsch$ for the
primary and $100\rsch$ for the secondary, where $\rsch$ is the Schwarzschild
radius. As found out in \cite{sun97}, changing the accretion radius only
affects the scaling of the result, as long as the radius is at or above
approximately $10\rsch$. Further, the number of particles escaping beyond $z >
4\stdev(z)$ vertically was counted, where $\stdev(z) = \pi z_0 / (2\sqrt{3})$,
the standard deviation of the $\sech^2(z/z_0)$ vertical distribution. These
particles represent such escapes from the disk that end up in the primary or
secondary jet as detailed in \cite{val06b}. The disk vertical scale height used
in the simulations was chosen as $z_0 = 260\usk\AU$ and the inner and
outer limits of the primary accretion disk were set to $3600\usk\AU$ and
$20600\usk\AU$ respectively.

\begin{figure}[htpb]
\plotone{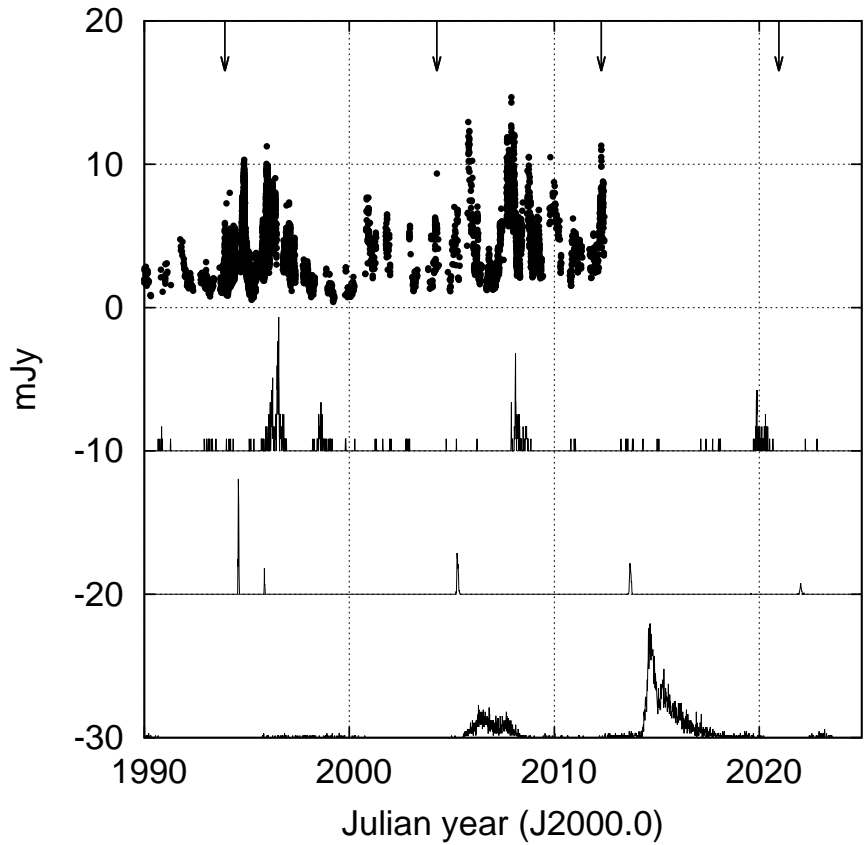}
\caption{Light curve since 1990 (\rvi{top panel}) and accretion counts of primary and
secondary black holes, and vertical escape counts (\rvi{three lower
panels}, arbitrary
scaling, offset by $-10$, $-20$ and $-30$, respectively).
The three historical and one predicted
precursor are indicated by arrows.}
\label{fig:data_vs_accre}
\end{figure}

The accretion and escape counts since 1990 for the final run with $5 \times 10^5$
particles and initial conditions from \cite{val10a} can be seen together with
light curve data in Figure~\ref{fig:data_vs_accre}. The major flares and
primary and secondary accretion line up well as expected from previous work
\citep{sun97}. The precursor flare timings occur definitely too early in
comparison with the particle flows displayed in
Figure~\ref{fig:data_vs_accre}. This applies to the
increases in vertical escapes of the particles from the primary disk (lowest
panel in Figure~\ref{fig:data_vs_accre}) as well as to accretion flows
to either of the black holes (the two middle panels in
Figure~\ref{fig:data_vs_accre}).

\begin{figure}[htpb]
\plotone{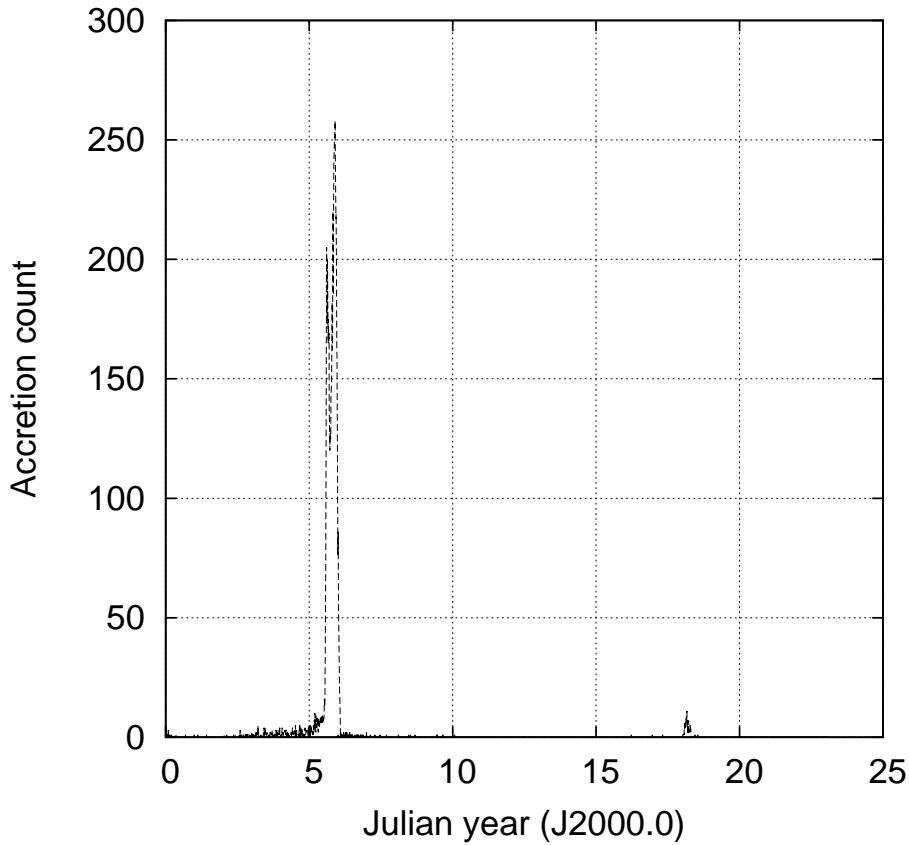}
\caption{Accretion counts of the primary (solid line) and the
secondary (dashed line) in a simulation of a circumsecondary disk with a
radius of $150\usk\AU$.}
\label{fig:sec_disk}
\end{figure}

\begin{figure}[htpb]
\plotone{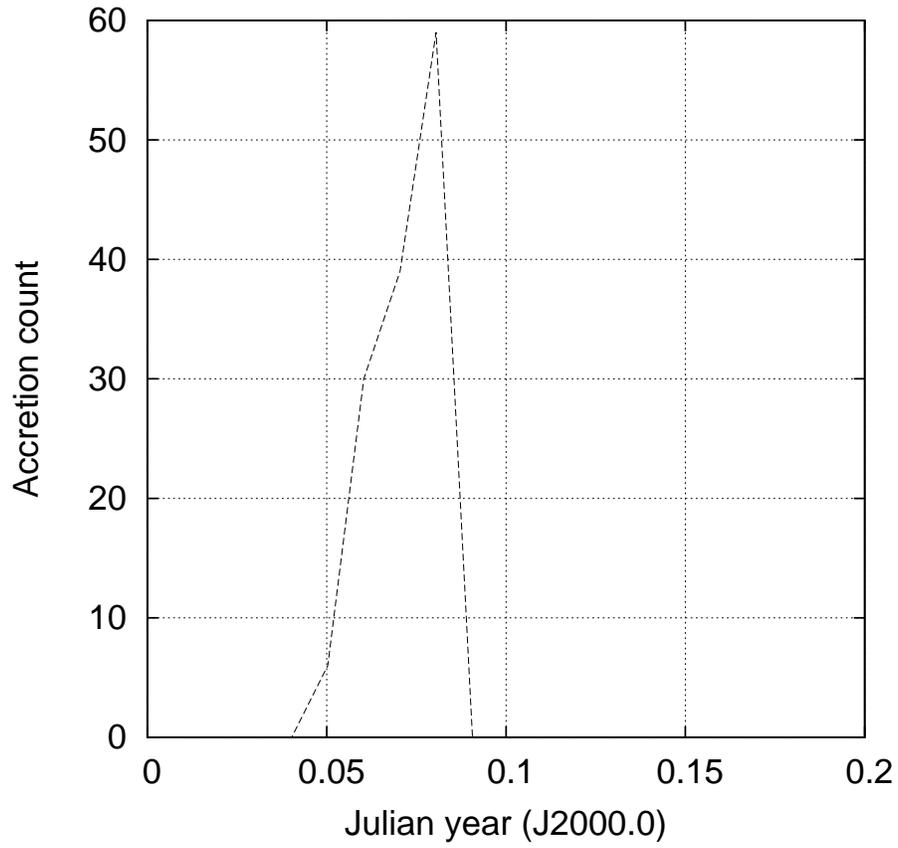}
\caption{Accretion counts of the
secondary (dashed line) in a simulation of a secondary colliding with a
spherical particle cloud of radius $300\usk\AU$ orbiting the primary.}
\label{fig:sec_coll}
\end{figure}

\rvi{%
To study the tidal effect of the primary on the secondary and the
proposed coronal gas cloud impacts in more detail, we used the code to
simulate the interaction between the secondary and a circumsecondary
disk and a cloud of particles representing a coronal cloud. Two
representative simulations were chosen, with both using initial
conditions from \cite{val10a} for the binary. For these simulations, 
the secondary accretion radius was set to $10\rsec$, 
where $\rsec$ is the Schwarzschild radius.
The interparticle gridded viscosity calculation was not used.
}

\rvi{%
In the first, we simulated a disk surrounding
the secondary to establish the tidal effect of the primary on the
accretion rate of the secondary. The disk was simulated with $10^4$
particles positioned in a constant planar density disk in the
$xy$-plane. The radius of the disk was set to $150\usk\AU$, comparable
to the tidally stable radius of $\sim50\usk\rsec\sim140\usk\AU$. This
configuration was simulated for approximately two secondary periods. 
The resulting accretion rates are illustrated in
Figure~\ref{fig:sec_disk}. The quiescent secondary accretion rate is greatly
enhanced by the tidal action of the primary during the close pericenter passage.
At this point, most of the disk is disturbed and accreted by
the secondary in a time scale of approximately $0.5\usk\yr$.
}

\rvi{%
In the second simulation, we used a spherical cloud of $10^4$ massive
particles, set at $z=4000\usk\AU$ with a keplerian rotation around the
primary, positioned so that it collides with the secondary. The radius
and total mass of the sphere were set to $300\usk\AU$ and $16\usk\Msun$,
respectively,
matching the theoretical predictions in the next section.
Figure~\ref{fig:sec_coll} graphs the secondary
accretion resulting from the collision. We find that the collision
induces a brief peak in the secondary accretion rate, with a timescale
of $\sim0.03\usk\yr$. This corresponds well with the theoretical
prediction in the next section.
}

\section{Discussion}

The simplest explanation of the above results is that the
secondary is impacting a thicker structure than the thin $\alpha$-disk used in
this simulation, and in much of the principal theoretical considerations and
simulations concerning OJ 287, such as \cite{iva98}. 

\begin{figure}[htpb]
\plotone{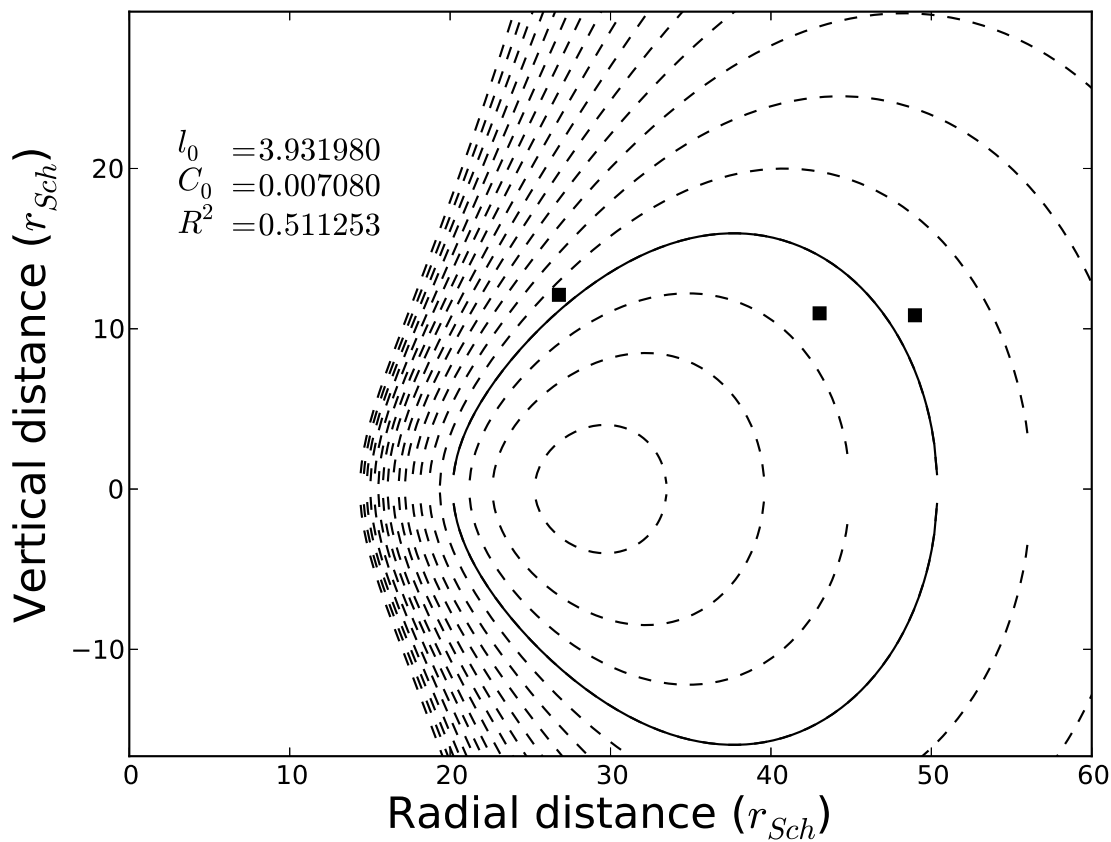}
\caption{Best fit of equation~\eqref{eq:equipressure} to model precursor
coordinates (thick line). The best fit values of $C_0$, and $l_0$ are
indicated, as is the goodness of the fit $R^2$. 
Additional equipressure surfaces with $C_0$ ranging
from $[0,2]$ times the best fit value (dashed lines) are also shown.}
\label{fig:equipressure}
\end{figure}

A plausible additional structure is a geometrically thick but optically thin
disk, such as the Polish doughnut model \citep{abr78,koz78,jar80,pac80}. 
In this model, the equipressure surfaces of the accretion disk can be
obtained by solving
\begin{equation}\label{eq:diffpolish}
\frac{\ud \theta}{\ud r} = 
-\frac{\partial_r g^{tt} - 2 l\partial_r g^{t\phi} + l^2 \partial_r g^{\phi\phi}}
{\partial_\theta g^{tt} - 2 l\partial_\theta g^{t\phi} + l^2 \partial_\theta g^{\phi\phi}},
\end{equation}
where $(t,r,\theta,\phi)$ are the usual Schwarzschild coordinates, 
$g^{\alpha\beta}$ are the components of the inverse of the metric
and $l$ is the specific angular momentum of a fluid element. Equation
\eqref{eq:diffpolish} can be
solved analytically for a Schwarzschild metric if a constant angular
momentum $l=l_0$ is assumed. The result is
\begin{equation}\label{eq:equipressure}
\frac{1}{\sin^2(\theta)} = -\frac{r^2[1 + 2C_0(1-r)]}{l^2(1-r)},
\end{equation}
where $r$ is in units of Schwarzschild radii and $C_0$ is the
integration constant enumerating the equipressure surfaces.
Figure~\ref{fig:equipressure}
shows a model fit of this equation to the model coordinates of the
precursor flares. Even though the fit describes a model with a nonrotating
primary, the change in the equipressure surfaces for a rotating primary is not
dramatic, as can be seen e.g. in \cite{qia09}.

Thus a reasonable possibility is a model with a cool, geometrically thin
accretion disk with a hot, geometrically thick corona. Additionally, the
periodical impacts of the secondary tear off gas clouds from the accretion
disk of the primary. Thus the corona of the primary's accretion disk is
expected to be partly composed of gas clouds, as in the model of \cite{sve94}.

In this model of precursor bursts we are observing a brightening of the
secondary black hole, not the primary. How can the secondary of mass $\secm$
exceed the brightness of the primary of mass $\prim$? The key is the relative
spin rate $\chi$ which is $\chi\sim0.28$ for the primary and, quite
reasonably, \rv{as will be shown,}
$\chi\sim1$ for the secondary. The jet brightness of an accreting black hole
with a jet is very sensitive to $\chi$ such that large $\chi$ systems are very
much brighter than low  $\chi$ systems of otherwise similar physical
parameters. The jets of the two black holes would be aligned, as the jet
direction is strongly influenced by the magnetic field of the primary accretion
disk \citep{pal10a,pal10b}. Thus both jets should have similar Doppler boosting
factors. 

We may additionally consider the secondary impact with a coronal gas
cloud in more detail. From \citep{leh96} we find the initial volume of
the gas clouds torn from the accretion disk of the primary by the
secondary during impacts to be
\begin{equation}
V_0 = \frac{\pi}{7} \eta^2 \rsec^2 h,
\end{equation}
where $\rsec$ is the Schwarzschild radius of the secondary black hole,
$h$ the primary accretion disk scale height and $\eta = (c/\vrel)^2$ is
a parameter of
interaction, while $c$
and $\vrel$ are the speed of light and
the velocity of the secondary relative to the disc, respectively. The
disk thickness is a function of the accretion rate $\dot{M}$ of the primary
black hole (Stella and Rosner 1984). This rate is not well known since
the observed emission of OJ 287 is jet-dominated, and the underlying disc-emission
is not directly observed. Based on the strength of the emission lines
\citep{sit85} and the observed (beamed) luminosity of
$\sim2\times10^{46}$ erg/s in the radio-UV range the value of disc
luminosity may be $\sim2\times10^{45}$ erg/s \citep{wor82} which
indicates $\dot{M}\sim0.8\times10^{-3}\medd$. $\medd$ is the accretion rate
of the primary that would give the Eddington luminosity for it. In the
following we will assume this accretion rate of the primary.

Further, $c/\vrel$ lies approximately in the  range of $4\text{--}12$;
for the impacts we are discussing here $c/\vrel\sim6$. Thus we may take
$\eta^2\sim1.3\times10^3$.  With $\rsec=4.2\times10^{13}$ cm we may
calculate $V_0$, and assuming that the initial cloud is spherical, its
radius is $R_0\sim15\usk\AU$.
In the disk model of \cite{sak81} with the scaling from
\cite{ste84}, we get the surface mass density of $\Sigma_0\sim10^{5}\usk
\gram\per\centi\meter^2$. The exact value depends on the viscosity
parameter $\alpha_g$ which is not known. Here we adopt $\alpha_g=0.5$
which may not be unreasonable \citep{kin07,ogi08,car12}. After
the initial impact, the cloud will expand by a factor of $k$. When the
secondary impacts this cloud again, the mass of the column interacting
with the secondary may be approximated as
\begin{equation}
\Delta m = \Sigma (\eta\rsec)^2 = \Sigma_0 \rsec^2\eta^2k^{-2},
\end{equation}
where $\Sigma\propto k^{-2}$ is the surface mass density of the gas
cloud after it has fully expanded and has reached the pressure balance
with the hot corona. We obtain $\Delta m \sim100
k^{-2}\usk\Msun$. The secondary crosses the extent of the cloud in a
timescale of $t_c \sim R_0k/\vrel \sim 1.5\times10^{-3}
k\usk\text{yr}$.
Here the difficulty lies in obtaining an estimate for the expansion
factor $k$ of the cloud. Following \cite{leh96} we may estimate it to
be at least $k=\tau^{4/7}\sim14$, where $\tau$ is the initial average optical
depth of the cloud, for at this point the cloud turns optically thin and
an outburst is observed. Thereafter the optical light curve follows
emission from an adiabatially expanding cloud; thus $k\sim20$ at the
time when the optical outburst is over. What happens thereafter is not
known from observations, but it is possible that the cloud cools rapidly
by radiation and may not expand very much again. We obtain the estimates
\begin{subequations}
\begin{gather}
\Delta m \sim 0.25 (k/20)^{-2}\usk\Msun  \\
t_c \sim 0.03 (k/20)\usk \text{yr} \\
\frac{\Delta m / t_c}{\medds} \sim 8 (k/20)^{-3}.
\end{gather}
\end{subequations}
Here $\medds\sim 1\usk\Msun/\yr$ is the Eddington accretion rate of
the secondary, assuming that the efficiency factor of the accretion
process $\epsilon=0.3$. \rv{This accretion rate gives the secondary the
luminosity of $\Ledd$.} \rvi{The timescale estimate is in agreement with
the simulation result obtained in the previous section for $k=20$.}

Most of the accreted matter will end up in a disk, and will be
transferred to the secondary black hole only slowly, in time which
exceeds the orbital period by orders of magnitude. We would expect the
immediate outburst only from matter which falls straight into the
\rv{innermost disk of the}
secondary, say, within \rv{$\sim3\rsec$} of the secondary
black hole.
This matter flow is only about \rv{$1/\eta\sim1/36$} of the flow
calculated above, and thus we expect \rv{outbursts} of roughly
$0.25(k/20)^{-3}\usk\Ledd$ during the
transit of the secondary through the gas cloud. The result scales with
$\alpha_g^{-0.8}$; thus it could not be much smaller than our estimate
but could possibly be bigger by a large factor, as $\alpha_g$ values
down to $\sim0.1$ have been reported \citep{kin07}. Thus, the secondary
could possibly accrete close to the Eddington rate $\medds$, which would
in turn indicate luminosity not so far from the Eddington value of
$\Ledd = 2 \times 10^{46}\usk\erg\per\second$. 

\rv{Here we have assumed
that the matter falling at or near the inner edge of the accretion disk
is transferred to the secondary black hole and its jet in a timescale
$\lesssim t_c$. This would not be the case in the standard steady
accretion process where the timescale is of the order of $\sim
1000\usk\yr$. However, it has been demonstrated that the accretion
timescale is much faster than the standard steady rate when we are
considering accretion of magnetic flux to the jet where the relevant
timescale for matter falling near $\risco$ is of order $\sim 10$ orbital
periods, or even less \citep{kro05}. For the secondary this is
equivalent to a timescale of $\sim7$ days. However, 
magnetohydrodynamic accretion is under intense study, and it is
difficult to be very specific about the time scale at this time
\citep{bec09}.
Fast timescales have also been demonstrated in the case
when a strong tidal perturbation applies to the disk
\citep{byr86,byr87,lin88,goo93}. 
In this case it is the tidal perturbation of the
primary on the secondary disk which causes accretion in the orbital
timescale of the inner disk, i.e. in a timescale of a few days.
}

\rv{
The last element of the model that needs to be considered in detail is
the spin of the secondary component. We propose that the binary model
naturally leads to a high present time spin value of the secondary.
First, we
note that the Sakimoto--Coroniti disk model is gravitationally unstable
at distances over $\sim 2\usk\parsec$ for
the OJ 287 primary \citep{sak81}. However, during the coalesence of
a supermassive black hole binary with a large mass ratio, the binary is
expected to spend a longer time at smaller separations, or that the
evolution would stall in what is known as the ''final parsec problem''
\citep{ber06}.
A recent study by \citet{iwa11} uses a theoretical prediction by
\citet{mat07} to calculate
that a coalescing supermassive black hole binary of a comparable mass
ratio to OJ 287 would reach a semi-major axis of $2\usk\parsec$
in $\sim5\times10^{7}\usk\yr$ starting from an initial semi-major axis of
$20\usk\parsec$. Based on numerical work they find the total merger
timescale to be greater than
$\sim2.5\times10^8\usk\yr$, for a system with parameters like OJ 287, 
which thus leaves at least
$2.0\times10^8\usk\yr$ for the secondary to accumulate spin through
accretion of matter from the accretion disk of the primary.
}

\rv{
The accretion spin-up timescale can be estimated from the results
initially derived in \citet{bar70} and further elaborated in
\citet{tho74}. As the estimate of the current mass of the secondary is 
$1.4\times 10^8\usk\Msun$, the amount of rest mass it must have accreted
to reach a spin of $\chi\sim1$ is $\sim1\times 10^8\usk\Msun$.
By the reasoning above and in \citet{leh96}, we would expect
a mass of $2\Sigma\eta\rsec^2$ to be accreted by the secondary
during one period $P$, which leads to a fiducial accretion rate of 
$\dot{M} = 2\Sigma\eta\rsec^2/P$. The scaling in \citet{ste84} gives
$\Sigma\propto r^{-3/5}$ at larger distances, and for the rest 
$\eta\propto P^2\propto r^3$. Using the values at $\Sigma_0=10^{5}\usk
\gram\per\centi\meter^2$, $\eta_0=36$ and $P_0=12\usk\yr$ at
$r=r_0=10^4\usk\AU$, we find 
$\dot{M} \sim 0.5(r/r_0)^{9/10}\usk\Msun/\yr$. Thus we would expect the
accretion rate to have been larger in the past, though considering the
growth of eccentricity documented in \cite{iwa11}, the impacts of the
secondary have been at a distance of $\sim r_0$ during most of the
orbital evolution. Thus using the value
$\dot{M} = 0.5\usk\Msun/\yr$ leads to an upper limit estimate of the 
spin-up timescale of $10^8\usk\Msun/\dot{M}\sim 2\times10^8\usk\yr$,
which is in line with the estimated time available for the accretion to
occur. Further, it is possible that the orbital plane of the secondary may have 
been closer the plane of the accretion disk of the primary before, having
attained the current value of inclination at $\sim90\degree$ at a later
time, as evidenced by the evolution of inclination in \citet{iwa11}.
This would have increased the accretion by the secondary, and further
shortened the spin-up timescale.
Based on these considerations we would expect the secondary in the binary black
hole model to have a high spin, at least compared to the primary spin
value. 
}

\rv{The resulting luminosity for the secondary is then} 
comparable to the maximum bremsstrahlung luminosity arising
from the hot bubbles torn off the disk \citep{iva98} or the luminosity
of the primary accreting at the rate of $\sim10^{-3}\medd$,
amplified by a reasonable forward beaming Doppler factor.
Since the primary and secondary are expected to have similar Doppler
boosting factors, precursor flares originating from the secondary black
hole could reasonably exceed the primary black hole luminosity for brief
periods.  As such, the case for precursor flares originating from the
secondary is quite plausible.

One of the best ways to verify the presence of the secondary jet in the
radiation of OJ 287 is to look for the shortest variability timescale.
It should be related to the orbital period of the innermost stable
circular orbit in the co-rotating case which is approximately 3.8 hours
for the secondary, and 100 days for the primary. Due to Doppler boosting
in the jet, variability time scales down to about 15 minutes may appear
in the secondary jet and about 5 days in the primary jet. Therefore
during secondary bursts the variability may be especially rapid. Some
evidence for this has been seen in the light curve of OJ 287 prior to
the 1983 great outburst \citep{val84}. 

Considering that the semi-major axis of the system corresponds to an
observed angular diameter of $\sim 0.01\usk\text{mas}$, it might be
possible to follow the orbital motion of the binary in the sky. This
requires observational capabilities to progress to a spatial resolution
at 10 microarcsecond level, possible with e.g. the GRAVITY instrument
\citep{eis11}. In addition to directly confirming the binary nature of
the system, this would allow explicit coordination of light curve events
with the phase of the binary orbit.

\section{Conclusions}

We have studied three precursor flares in the light curve of OJ 287. We
propose a model where these outbursts are caused by the secondary black
hole impacting on a thick disk of the primary before the actual
impact on the primary accretion disk. The precursor outburst is
triggered by gas falling into the secondary
black hole leading to its jet brightening.

Using this model we predict a new precursor flare of to occur around
$\nextpc$, \rv{with a brightness of approximately $2\usk\magn$ above the
quiescent level.} It would be especially interesting to look for intra-day
variability during this outburst. Also we expect that intra-day
variability will appear prominently in smaller outburst events between
now and September 2013 when the secondary hits the primary accretion
disk, and moves to the far side of the disk as seen by us. 

\acknowledgments
P. Pihajoki acknowledges the support of Turku University Foundation 
(grant no.~7642) and Magnus Ehrnrooth foundation (grant no.~Ta2012n6).

This work has been partly supported by the Polish MNiSW grant under the 
contract No. 3812/B/H03/2009/36.

\bibliographystyle{apj}

\begin{thebibliography}{}
\bibitem[Abramowicz et al. (1978)]{abr78} Abramowicz, M., Jaroszynski, M. \& Sikora, M. 1978, \aap, 63, 221
\bibitem[Anupama et al. (1994)]{anu94} Anupama, G.C., Kembhavi, A.K., Prabhu, T.P., Singh, K.P. \& Bhat, P.N. 1994, \aas, 103, 315
\bibitem[Baliyan et al. (2005)]{bal05} Baliyan, K.S., Joshi U.C. \& Ganesh S. 2005, BASI, 33, 181
\bibitem[Bardeen (1970)]{bar70} Bardeen, J.M. 1970, Nature, 226, 64
\bibitem[Beckwith et al. (2009)]{bec09} Beckwith, K., Hawley, J.F. \& Krolik, J.H. 2009, \apj, 707, 428
\bibitem[Berczik et al. (2006)]{ber06} Berczik, P., Merritt, D., Spurzem, R. \& Bischof, H. 2006, \apj, 642, L21
\bibitem[Budyn et al. (2010)]{bud10} Budyn, M., Zola, S. \& Wojcik, K. 2010, ASPC, 435, 87
\bibitem[Byrd et al. (1986)]{byr86} Byrd, G.G., Valtonen, M.J., Valtaoja, L. \& Sundelius, B. 1986, \aap, 166, 75
\bibitem[Byrd et al. (1987)]{byr87} Byrd, G.G., Sundelius, B. \& Valtonen, M. 1987, \aap, 171, 16
\bibitem[Carciofi et al. (2012)]{car12} Carciofi, A.C. et al. 2012, \apj, 744, L15
\bibitem[Chandra et al. (2011)]{cha11} Chandra, S., Baliyan, K.S.,  Ganesh, S. \& Joshi, U.C. 2011, \apj, 731, 118
\bibitem[Eisenhauer et al. (2011)]{eis11} Eisenhauer, F. et al. 2011, The Messenger, 143, 16
\bibitem[Fiorucci et al. (1996)]{fio96} Fiorucci, M. \& Tosti, G. 1996, \aas, 116, 403
\bibitem[Gezari et al. (2012)]{gez12} Gezari, S. et al. 2012, Nature, 485, 217
\bibitem[Gonzalez-Perez et al. (2001)]{gon01} Gonzalez-Perez J.N.,
Kidger M.R. \& Martin-Luis F. 2001, AJ, 122, 2055
\bibitem[Goodman, J. (1993)]{goo93} Goodman, J. 1993, \apj, 406, 596
\bibitem[Hawley and Krolik (2006)]{haw06} Hawley, J.F. \& Krolik, J.H. 2006, \apj, 641, 103
\bibitem[Ivanov et al. (1998)]{iva98} Ivanov, P.B., Igumenschev, I.V. \& Novikov, I.D. 1998, \apj, 507, 131
\bibitem[Iwasawa et al. (2011)]{iwa11} Iwasawa, M. et al. 2011, \apj, 731, L9
\bibitem[Jaroszynski et al. (1980)]{jar80} Jaroszynski, M., Abramowicz, M.A. \& Paczy\'nski, B. 1980, \actaa, 30, 1

\bibitem[Jester et al (2005)]{jes05} Jester, S. et al. 2005, \aj, 130, 873

\bibitem[Kidger and Takalo (1993)]{kid93} Kidger, M.R., \& Takalo, L.O. 1993, IAU Circ., 5909, 1
\bibitem[Kidger et al. (1994)]{kid94} Kidger, M.R. et al. 1994, Intensive Monitoring of OJ287, eds. M.R. Kidger \& L.O. Takalo, Tuorla Observatory Rep. 174, 106 
\bibitem[Kidger et al. (1995)]{kid95} Kidger, M.R. et al. 1995, \aas, 113, 431
\bibitem[King et al. (2007)]{kin07} King, A.R., Pringle, J.E. \& Livio, M. 2007, \mnras, 376, 1740 
\bibitem[Kozlowski et al. (1978)]{koz78} Kozlowski, M., Jaroszynski, M. \& Abramowicz, M.A. 1978, \aap, 63, 209
\bibitem[Krolik et al. (2005)]{kro05} Krolik, J.H., Hawley, J.F. \& Hirose, S. 2005, \apj, 622, 1008

\bibitem[Landolt (2009)]{lan09} Landolt A.U. 2009, AJ, 137, 4186
\bibitem[Lehto \& Valtonen (1996)]{leh96} Lehto, H.J., \& Valtonen, M.J. 1996, \apj, 460, 207
\bibitem[Lin et al. (1988)]{lin88} Lin, D.N.C., Pringle, J.E. \& Rees, M.J. 1988, \apj, 328, 103
\bibitem[Matsubayashi et al. (2007)]{mat07} Matsubayashi, T., Makino, J.  \& Ebisuzaki, T. 2007, \apj, 656, 879
\bibitem[Miller (1976)]{mil76} Miller, R.H. 1976, J.Comput.Phys., 21, 400
\bibitem[Neilsen et al. (2011)]{nei11} Neilsen, D. et al. 2011, Proc. Nat. Acad. Sci., 108, 12641
\bibitem[Ogilvie \& Dubus (2008)]{ogi08} Ogilvie, G.I. \& Dubus, G. 2008, \mnras, 320, 485
\bibitem[Paczy\'nski and Wiita (1980)]{pac80} Paczy\'nski, B. \& Wiita, P.J. 1980, \aap, 88, 23
\bibitem[Palenzuela et al. (2010a)]{pal10a} Palenzuela, C., Garrett, T., Lehner, L. \& Liebling, S.L. 2010a, Phys.Rev.D, 82, 44045
\bibitem[Palenzuela et al. (2010b)]{pal10b} Palenzuela, C., Lehner, L. \& Liebling, S.L. 2010b, Science, 329, 927
\bibitem[Qian et al. (2009)]{qia09} Qian, L. et al. 2009, \aap, 498, 471
\bibitem[Sagar et al.(2004)]{sag04} Sagar, R., Stalin, C.S., Gopal-Krishna \& Wiita, P.J. 2004, \mnras, 348, 176
\bibitem[Sakimoto \& Coroniti (1981)]{sak81} Sakimoto, P.J., \& Coroniti, F.V. 1981, \apj, 247, 19
\bibitem[Shakura \& Sunyaev (1973)]{sha73} Shakura, N.I., \& Sunyaev, R.A. 1973, \aap, 24, 337
\bibitem[Sillanp\"a\"a et al.(1988)]{sil88} Sillanp\"a\"a, A., Haarala, S., Valtonen, M.J., Sundelius, B. \& Byrd,     G.G.    1988, \apj, 325, 628
\bibitem[Sillanp\"a\"a et al.(1996a)]{sil96a} Sillanp\"a\"a et al.  1996a, \aap, 305, L17
\bibitem[Sillanp\"a\"a et al.(1996b)]{sil96b} Sillanp\"a\"a et al.  1996b, \aap, 315, L13
\bibitem[Sitko \& Junkkarinen (1985)]{sit85} Sitko, M.L. \& Junkkarinen, V.T. 1985, \pasp, 97, 1158
\bibitem[Stella \& Rosner (1984)]{ste84} Stella, L., \& Rosner, R. 1984, \apj, 277, 312
\bibitem[Thorne (1974)]{tho74} Thorne, K.S., 1974, \apj, 191, 507
\bibitem[Sundelius et al. (1996)]{sun96} Sundelius, B., Wahde, M., Lehto, H.J. \& Valtonen, M.J. 1996, Blazar Continuum Variability, ASP Conf. Ser., 110, 99
\bibitem[Sundelius et al. (1997)]{sun97} Sundelius, B., Wahde, M., Lehto, H.J. \& Valtonen, M.J. 1997, \apj, 484, 180
\bibitem[Svensson \& Zdziarski (1994)]{sve94} Svensson, R., \& Zdziarski, A.A. 1994, \apj, 436, 599
\bibitem[Valtaoja et al. (1984)]{val84} Valtaoja, E. et al. 1984, Nature, 314, 148 
\bibitem[Valtonen \& Lehto (1997)]{val97}  Valtonen, M.J. \& Lehto, H.J. 1997, \apj, 481, L5
\bibitem[Valtonen et al. (2006a)]{val06a}  Valtonen, M.J. et al. 2006a, \apj, 643, L9
\bibitem[Valtonen et al. (2006b)]{val06b}  Valtonen, M.J. et al. 2006b, \apj, 646, 36
\bibitem[Valtonen (2007)]{val07}  Valtonen, M.J. 2007, \apj, 659, 1074
\bibitem[Valtonen et al. (2008a)]{val08a}  Valtonen, M.J., Kidger, M., Lehto, H. \& Poyner, G. 2008a, \aap, 477, 407

\bibitem[Valtonen et al. (2008b)]{val08b}  Valtonen, M.J. et al. 2008b, Nature, 452, 851



\bibitem[Valtonen et al. (2009)]{val09}  Valtonen, M.J. et al. 2009, \apj, 698, 781
\bibitem[Valtonen et al. (2010a)]{val10a}  Valtonen, M.J. et al. 2010a, \apj, 709, 725
\bibitem[Valtonen et al. (2010b)]{val10b}  Valtonen, M.J. et al. 2010b, Cel. Mech. Dyn. Astr., 106, 235


\bibitem[Valtonen et al.(2011)]{val11}  Valtonen, M.J. et al. 2011, \apj, 742, 22
\bibitem[Worrall et al.(1982)]{wor82}  Worrall, D.M. et al. 1982, \apj, 261, 403
\end{thebibliography}

\end{document}